\documentclass[aps,amsmath,amssymb, prapplied, twocolumn,superscriptaddress]{revtex4-2}

\usepackage{comment} 

\usepackage{graphicx}
\usepackage{dcolumn}
\usepackage{tikz}
\usepackage{bm}
\usepackage[normalem]{ulem} 
\usepackage{multirow}

\makeatletter
    \renewcommand\@make@capt@title[2]{%
     \@ifx@empty\float@link{\@firstofone}{\expandafter\href\expandafter{\float@link}}%
      {#1}\@caption@fignum@sep#2\quad}%
    \makeatother
   
\makeatletter 
\renewcommand{\fnum@figure}{FIG.~\thefigure}
\makeatother

\newcommand{\sz}[0]{\sigma^z}
\newcommand{\sx}[0]{\sigma^x}

\newcommand{\ket}[1]{|#1\rangle}
\newcommand{\bra}[1]{\langle#1|}

\begin{document}

\title{A scalable 2-local architecture for quantum annealing of Ising models with arbitrary dimensions}

\author{Ana Palacios}
\affiliation{%
 Qilimanjaro Quantum Tech, Carrer de Veneçuela, 74, Sant Martí, 08019, Barcelona, Spain}
\affiliation{Departament de F\'{i}sica Qu\`{a}ntica i Astrof\'{i}sica, Facultat de F\'{i}sica,
Universitat de Barcelona, 08028 Barcelona, Spain}
\affiliation{Institut de Ci\`{e}ncies del Cosmos, Universitat de Barcelona,
ICCUB, Mart\'{i} i Franqu\`{e}s 1, 08028 Barcelona, Spain.}
\author{Artur Garcia-Saez}
\affiliation{%
Qilimanjaro Quantum Tech, Carrer de Veneçuela, 74, Sant Martí, 08019, Barcelona, Spain}
\affiliation{Barcelona Supercomputing Center, Plaça d'Eusebi Güell, 1-3, Les Corts, 08034 Barcelona, Spain}
\author{Bruno Juliá-Díaz}
\affiliation{Departament de F\'{i}sica Qu\`{a}ntica i Astrof\'{i}sica, Facultat de F\'{i}sica,
Universitat de Barcelona, 08028 Barcelona, Spain}
\affiliation{Institut de Ci\`{e}ncies del Cosmos, Universitat de Barcelona,
ICCUB, Mart\'{i} i Franqu\`{e}s 1, 08028 Barcelona, Spain.}

\author{Marta P. Estarellas}%
\affiliation{%
Qilimanjaro Quantum Tech, Carrer de Veneçuela, 74, Sant Martí, 08019, Barcelona, Spain}

\date{\today}

\begin{abstract}
Achieving densely connected hardware graphs is a challenge for most quantum computing platforms today, and a particularly crucial one for the case of quantum annealing applications. In this context, we present a scalable architecture for quantum annealers to realize effective Ising Hamiltonians of arbitrary connectivity. Our proposal 
consists on a resource-efficient configuration based on a hardware graph where physical qubits are connected to at most other 3 and containing exclusively 2-local interactions. We derive this configuration based on chains of qubits encoding logical variables by describing the problem graph in terms of triangles. We thus present a promising new route to scale up devices dedicated to classical optimization tasks within the quantum annealing paradigm.
\end{abstract}
\maketitle


\section{Introduction} 
Quantum annealing (QA) \cite{kadowaki_quantum_1998, farhi_quantum_2000, albash_adiabatic_2018} is an analog form of quantum computation which encodes the solution to a certain problem in the low-energy eigenspace of a final Hamiltonian, $H_f$. This is achieved by preparing the ground state of a simple Hamiltonian, $H_0$, and interpolating towards $H_f$ following
\begin{equation}
    H(t) = (1 - \lambda(t)) H_0 + \lambda(t) H_f
    \label{eq:annealing} \,,
\end{equation}
where $\lambda(0) = 0$ and $\lambda(t_f) = 1$ with $t_f$ being the total time of the anneal. In the context of optimization, $H_f$ encodes a function that one intends to minimize, i.e., a cost function, and thus obtaining the ground state of $H_f$ (or low-lying excited states) provides with optimal solutions to the relevant minimization problem.

An interesting Hamiltonian in the context of annealing applications is the Ising model
\begin{equation}
    H_{\text{ising}} = H_f = \sum_i h_i \sz_i + \sum_{j>i} J_{ij} \sz_i \sz_j 
    \label{eq:general_ising} \,,
\end{equation}
where the $\{h_i\}$ are local fields, the $\{J_{ij}\}$ are the couplings (which determine the connectivity of the problem) and $\sz_i$ is the Pauli-$z$ matrix of the $i$-th qubit.
This Hamiltonian can be implemented in a number of quantum computing platforms, such as superconducting quantum 
circuits~\cite{weber_coherent_2017, krantz_quantum_2019} or neutral 
atoms~\cite{pohl_dynamical_2010, de_leseleuc_accurate_2018}. The cost function of many relevant classical optimization problems can be cast in the shape of Eq.~\eqref{eq:general_ising}, often requiring all-to-all interactions \cite{lucas_ising_2014}. However, establishing a dense connectivity among all qubits remains a challenging task; e.g., in superconducting platforms the hard-wiring of all the necessary links is not scalable due to crosstalk and packing issues within the chip. Thus, the original (or logical) problem defined on $N$ logical variables needs to be encoded in a larger space, defined on $n$ physical variables ($n>N$), in such a way that we recover the logical problem by adding restrictions on the larger space. 

These mappings, referred to as embeddings, have been studied in the literature. Based on~\cite{kaminsky_scalable_2004}, it was proposed in~\cite{choi_minor-embedding_2008, choi_minor-embedding_2011} to make up for the required connectivity by establishing chains of qubits that redundantly represent the same logical variable. This is then implemented in a fixed family of graphs within which one searches for an optimal embedding, an approach known as minor embedding. Another alternative is the Lechner-Hauke-Zoller (LHZ) scheme~\cite{lechner_quantum_2015}, which entails switching to the description of the problem in terms of parities between the original variables. This allows to encode the original cost function into single-qubit terms at the expense of additional $3-$ and $4-$body interactions (or 2-local interactions mediated by qutrits). The latter are hard to achieve experimentally, despite some early proposals~\cite{puri_quantum_2017, schondorf_nonpairwise_2019} and prototypes~\cite{ menke_demonstration_2022}.
A slight modification of LHZ was proposed in \cite{rocchetto_stabilizers_2016} that allows to implement the parity encoding using only 2-local terms (and 2-level systems) by means of odd parity constraints. This approach requires $N^2$ qubits, antiferromagnetic constraints and a graph where parity qubits require connection to 12 other qubits and ancilla qubits to other 4. The implementation of this scheme was studied on D-Wave devices in~\cite{cattelan_scalable_2024}.
Recently, a new method based on perturbative gadgets has been proposed for realizing all-to-all connectivity~\cite{mozgunov_precision_2023}, but at the cost of interaction strengths of $\mathcal{O}(N^6)$ for an $N$-qubit problem.

In this work we provide an architecture that enables encoding arbitrary Ising models (schemed in Fig.~\ref{fig:full_scheme_N5}).  It requires only $N(N-2)$ qubits, ferromagnetic constraints and 2-local interactions. Remarkably, in our proposal every qubit is linked to other three physical qubits at most. We develop this architecture step by step from a decomposition of the original problem in terms of triangles of logical variables. Our construction naturally performs a change of variables of the original logical problem, which has implications for QA on the logical level further discussed in Appendix~\ref{sec:factorised_encoding}. A different discussion constrained to this change of logical variables is found in \cite{fujii_energy_2022, fujii_eigenvalue-invariant_2023}.


\begin{figure}
    \centering
    \includegraphics[width=0.3\textwidth]{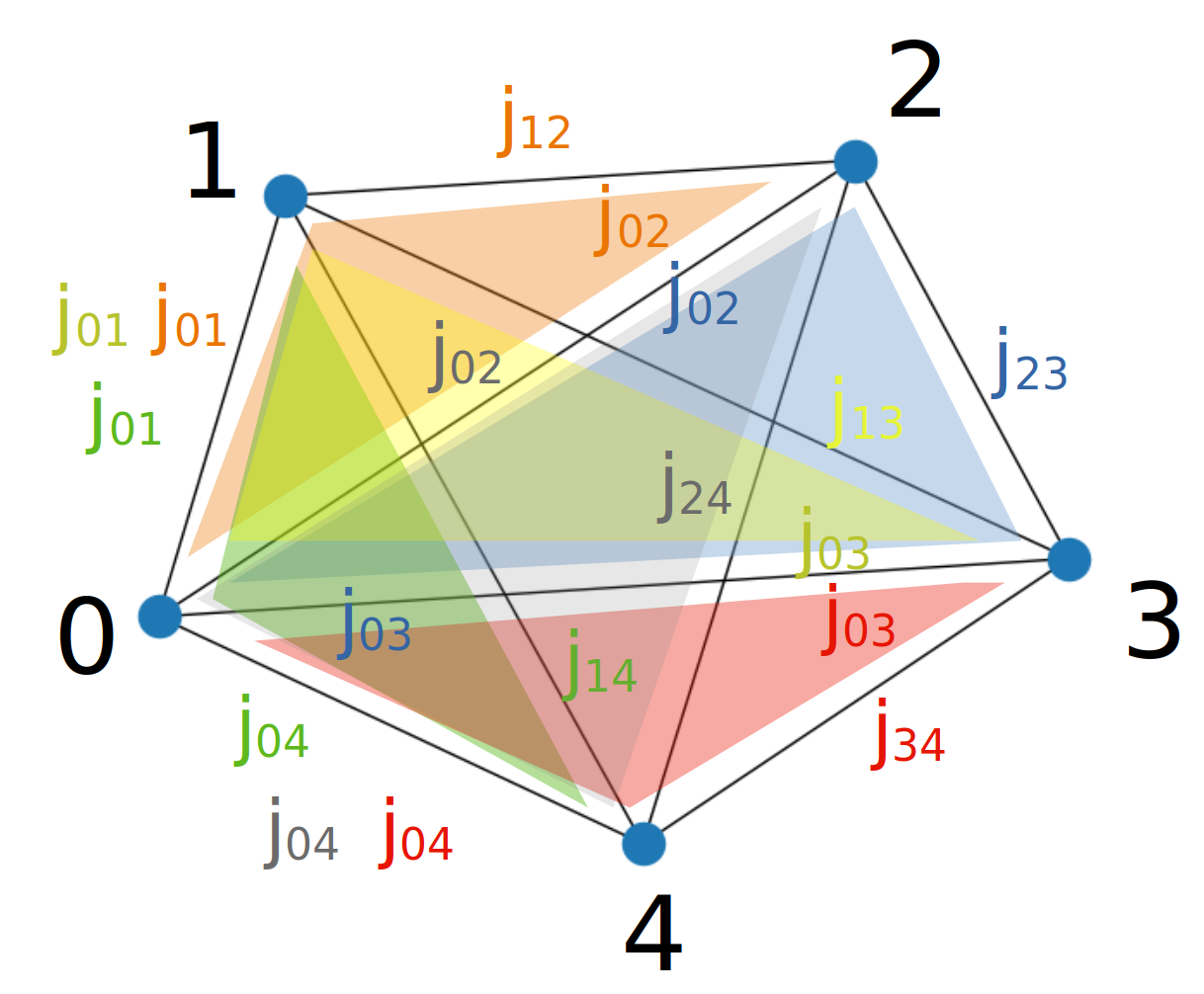}
    \caption{Possible partition of the cost Eq.~\eqref{eq:ising_Z2} for a fully connected graph of $N=5$, into that of individual triangles. The couplings $\{J_{kl}\}$ here are implicitly indicated by the gray links, while the lower-case $\{j_{kl}\}$ refer to the single-triangle cost in Eq.~\eqref{eq:logical_ham_single_cell}. Note that the different colors in this illustration play the role of the superscript $r$ in~\eqref{eq:logical_ham_single_cell}.}
    \label{fig:triangle_partition_costfun}
\end{figure}

\section{Triangle decomposition of the Ising problem}\label{sec:triangle_decomposition}
Let us first consider the less general class of $\mathbb{Z}_2$-symmetric classical Ising models for a system of size $N$, described as:
\begin{gather}
    H_{\mathbb{Z}_2} = \sum_{l>k} J_{kl} \sz_k \sz_l  \,.
    \label{eq:ising_Z2}
\end{gather}
This all-to-all connected Hamiltonian can be 
subdivided as the sum of the Hamiltonians of a set of individual triangles, $\mathcal{T} = \{r\}$:
%
\begin{gather}
    H_{\mathbb{Z}_2} = \sum_{r \in \mathcal{T}} H_{r} \,, \\
    H_{r} = 
    j^{r}_{kl} \sz_k \sz_l + 
    j^{r}_{km} \sz_k \sz_m + 
    j^{r}_{lm} \sz_l \sz_m \,.
    \label{eq:logical_ham_single_cell}
\end{gather}
In the previous expression, $r$ indexes a logical triangle containing nodes $\{k, l, m\}$ and $j^{r}_{kl}$ are the corresponding couplings. Note that the choice of $\mathcal{T}$ is not unique. The equality holds as long as $J_{kl} = \sum_{r'} j^{r'}_{kl}$, where $r'$ runs through all triangles containing edge $(k, l)$. An illustration of such a partition for a problem graph of $N=5$ is presented in Fig.~\ref{fig:triangle_partition_costfun}.
\begin{table}[]
    \hspace{0.02cm}
    a) \quad
    \begin{tabular}{|c|cl|cl|cl|cl|}
    \hline
    $\eta,  \xi $    & \multicolumn{2}{c|}{$00$}         & \multicolumn{2}{c|}{$01$}         & \multicolumn{2}{c|}{$10$}         & \multicolumn{2}{c|}{$11$}         \\ \hline
    $k, l, m$ & \multicolumn{2}{c|}{$000$, $111$} & \multicolumn{2}{c|}{$110$, $001$} & \multicolumn{2}{c|}{$011$, $100$} & \multicolumn{2}{c|}{$101$, $010$} \\ \hline
    \end{tabular}
    \label{tab:Z2_configs}
    \vspace{0.2cm}
    \\ b) \quad
    \begin{tabular}{|c|cc|cc|cc|cc|}
    \hline
    $\eta, \xi $    & \multicolumn{2}{c|}{$00$}          & \multicolumn{2}{c|}{$01$}          & \multicolumn{2}{c|}{$10$}          & \multicolumn{2}{c|}{$11$}          \\ \hline
    $\gamma $        & \multicolumn{1}{c|}{$0$}   & $1$   & \multicolumn{1}{c|}{$0$}   & $1$   & \multicolumn{1}{c|}{$0$}   & $1$   & \multicolumn{1}{c|}{$0$}   & $1$   \\ \hline
    $k, l, m$ & \multicolumn{1}{c|}{$000$} & $111$ & \multicolumn{1}{c|}{$110$} & $001$ & \multicolumn{1}{c|}{$011$} & $100$ & \multicolumn{1}{c|}{$101$} & $010$ \\ \hline
    \end{tabular}
    \caption{Encoding of logical triangle configurations of nodes $\{k, l, m\}$ in physical qubits $\{\eta, \xi\}$ for the: (a) $\mathbb{Z}_2$-symmetric case, and (b) non-symmetric case including local fields. 
    The latter includes an additional sign qubit, such that the logical triangle formed by $\{k, l, m\}$ is now described by physical qubits $\{\eta, \xi, \gamma\}$.}
    \label{tab:frust_configs}
\end{table}
%

Now we explain how we can encode each logical triangle in terms of physical qubits. In fact, we just need two physical qubits to describe the energetically distinct states of the model due to inversion symmetry. A map between the logical configuration in a triangle and the corresponding configuration of the physical qubits that describe it is illustrated in Table~\ref{tab:frust_configs}a. There, without loss of generality, we adopt the convention of always labeling the all-parallel configuration as ``00''.

We can determine the Hamiltonian $H^{r}_{\text{physical}}$ acting on a physical qubit pair $\{\eta_r, \xi_r\}$ that reproduces the correct energies of triangle $r=\{k, l, m\}$ by solving a simple $4 \times 4$ linear system (see Appendix~\ref{sec:Hphys_cell})
. Following Table~\ref{tab:frust_configs}a, we find,
\begin{gather}
     H^{r}_{\text{physical}} = \Tilde{h}_{\eta_r} \sz_{\eta_r} + \Tilde{h}_{\xi_r} \sz_{\xi_r} + \Tilde{J}_{\eta_r \xi_r} \sz_{\eta_r} \sz_{\xi_r}
\end{gather}
where $\Tilde{h}_{\eta_r} = j_{kl}, \ \ \Tilde{h}_{\xi_r} = j_{km}$ and $\ \Tilde{J}_{\eta_r \xi_r} = j_{lm}$.

\subsection{Consistency conditions among triangles}\label{sec:consistency_conditions}
In order to ensure that the separate physical representations of the triangles refer to a logical state in the original model, we need to enforce some constraints. In particular, shared edges among triangles must be consistent with each other. Thus, if the configuration of one triangle is such that the shared edge holds parallel spins, this must also be the case for other triangular cells (also simply called cells from now on) that contain the same edge. One can check that for the encoding illustrated in Table.~\ref{tab:frust_configs}a., this can always be achieved by means of strong 2-local ferromagnetic interactions, as long as every cell has at least one edge that is not shared with any other cell. This non-shared edge is labeled with the ``11'' configuration in order to avoid the need for many-body terms.

\begin{figure}
    \centering
        \includegraphics[width=0.5\textwidth]{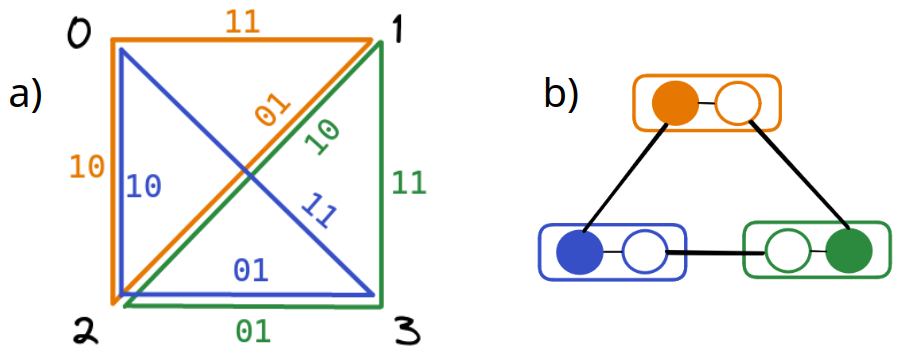}
    \caption{(a) Logical model of a 4-clique and (b) physical implementation according to the triangle mapping in Eq.~\eqref{eq:hsim_4clique}. The labels along the edges in a) correspond to the configuration of the qubits describing each triangle, indicating that said edge holds parallel spins while the opposite node is antiparallel. In b), empty and solid circles illustrate the qubits labeled 0 and 1 within the description of each triangle, respectively, and thicker lines indicate the strong ferromagnetic links imposing the constraints.}
    \label{fig:minimal_example_4clique}
\end{figure}

Let us illustrate this more clearly with the minimal example of a 4-clique, depicted in Fig.~\ref{fig:minimal_example_4clique}a. 
The orange and blue triangles have the (0, 2) edge in common, and thus the configurations they point to must be consistent along this edge. We will refer to the physical qubits describing the orange triangle as $o_0, o_1$ and to those describing the blue triangle as $b_0, b_1$. Note that we have made a small change of notation of the physical qubits to improve the readability of this example; the pair $(o_0, o_1)$ would correspond to $(\eta_o, \xi_o)$ in the notation we previously defined. If $o_0, o_1$  are in configuration 00 (all spins parallel) or 10 (the spins (0, 2) parallel and spin 1 antiparallel to the two), then $b_0, b_1$ must necessarily be in either 00 or 10. If we now look at the consistency condition between $o_0, o_1$ and the green triangle, described by $g_0, g_1$, we see that $(o_0, o_1) = \{00, 01\}$ can only happen if $(g_0, g_1)=\{00, 10\}$. The condition between blue and green triangles can be similarly deduced; we present a collection of all the conditions in Table~\ref{tab:constraints_4clique}. From this table it becomes apparent that we can enforce these constraints with 2-local interactions: we simply need to strongly couple ferromagnetically the pairs $(o_1, b_1), (o_0, g_1)$ and $(b_0, g_0)$. Notice that the key to get away with 2-local constraints is to choose the non-shared edge of each triangle to hold the 11 label, since linking the \{00, 11\} states to any other pair does require many-body interactions. Thus, the full Hamiltonian that simulates the $N=4$ clique is
\begin{gather}
    H_{p}^{4-clique} = H^{(o)}_\text{phys} + H^{(b)}_\text{phys} + H^{(g)}_{\text{phys}} + H_{constraints} 
    \label{eq:hsim_4clique} \\
    H_{constraints} = J_P (I - \sz_{o_1} \sz_{b_1}) + J_P (I - \sz_{o_0} \sz_{g_1}) + \nonumber \\
    + J_P (I - \sz_{b_1} \sz_{g_0}) \\
    H^{(i)}_{\text{phys}} = h^{(i)}_0 \sz_{i_0} + h^{(i)}_1 \sz_{i_1} + J^{(i)} \sz_{i_0} \sz_{i_1} \quad i = \{o, b, g\}
\end{gather}
where $J_P$ must be much larger than the characteristic energy scale of $H_{\mathbb{Z}_2}$ and the terms proportional to the identity were added to preserve the exact same energies as in $H_{\mathbb{Z}_2}$. Under these conditions, the Hamiltonian \eqref{eq:hsim_4clique} exactly encodes the spectrum of $H_{\mathbb{Z}_2}$ in its low-energy eigenspace. The corresponding graph of the physical implementation is illustrated in Fig.~\ref{fig:minimal_example_4clique}b.
\begin{table}[h]
    \centering
    \begin{tabular}{|c|c|c|c|c|c|c|c|}
    \cline{0-1} \cline{4-5} \cline{7-8}
        $o_0 o_1$ & $b_0 b_1$ && $o_0 o_1$ & $g_0 g_1$ & & $b_0 b_1$ & $g_0 g_1$\\ 
        \cline{0-1}  \cline{0-1} \cline{4-5} \cline{4-5} \cline{7-8} \cline{7-8}
        00 & 00 & \hspace{0.5cm} &  00 & 00 & \hspace{0.5cm}  &  00 & 00 \\
        10 & 10 & &  01 & 10 & &  01 & 01 \\ \cline{0-1} \cline{4-5} \cline{7-8}
        01 & 01 & &  10 & 01 & &  10 & 10 \\
        11 & 11 & &  11 & 11 & &  11 & 11 \\ \cline{0-1} \cline{4-5} \cline{7-8}
    \end{tabular}
    \caption{Compilation of the consistency conditions that ensure that the configuration of $\{o_0, o_1, b_0, b_1, g_0, g_1\}$ corresponds to a physical configuration of spins $\{0, 1, 2, 3\}$. Configurations on the same row are compatible with each other.
    }
    \label{tab:constraints_4clique}
\end{table}

Thus, back in the general case, the full physical Hamiltonian embedding the logical problem is
\begin{gather}
    H_p = \sum_{\text{all }r} H^{r}_{\text{physical}} + H_{\text{constraints}}
    \label{eq:shape_physical_hamiltonian} \,, \\
    H_{\text{constraints}} = 
    \sum_{r, s / r \cap s \neq \emptyset} 
    J_P (I - \sz_{\omega_r} \sz_{\omega^\prime_s}) \,,
\end{gather}
where $\omega$ and $\omega^\prime$ refer to either $\eta$ or $\xi$ depending on the internal labeling of triangle $r$.
For sufficiently strong ferromagnetic penalties $J_P$, the Hamiltonian~\eqref{eq:shape_physical_hamiltonian} exactly encodes the spectrum of the original $H_{\mathbb{Z}_2}$ in its low-energy eigenspace. 

We note that the generation of strong enough ferromagnetic interactions as system size grows is an issue that presents itself in all penalty-enforced embeddings. Some studies on the scalability of penalties have been carried out for the minor embedding and LHZ cases. 
They show linear \cite{venturelli_quantum_2015, lanthaler_minimal_2021} and up to quadratic \cite{lanthaler_minimal_2021} penalty growth in the range of problems and system sizes examined; similar results are expected for the present scheme due to its similarities with them. We tackle this limitation further in this work by proposing driver Hamiltonians that ease these penalty strength requirements.

\subsection{Scaling up the construction}\label{sec:scaling_up}
For an arbitrary graph of size $N$ we can always find some family of decompositions in terms of triangles such that every edge is accounted for at least once, and such that all triangles hold a non-shared edge. This is achieved by choosing all the triangles in $\mathcal{T}$ such that they have a selected node $k^*$ in common. Starting from a single triangular cell, the graph can be increased in an inductive way as follows:
for the addition of each new node $i = \kappa+1$, we include the triangles that complete the missing links to restore full connectivity by selecting the triads $\{i, k, k^*\}$, where $k=1, ..., k^*-1, k^*+1, ..., \kappa$ is one of the remaining former nodes.
With this scheme, the addition of an $\kappa$-th qubit adds $\kappa-2$ new triangles, and thus $2(\kappa-2)$ physical qubits, such that the total number of qubits $n$ required by the physical implementation is:
\begin{equation}
    n = 2 \sum_{\kappa=1}^{N-2} \kappa =  (N-1)(N-2)
\end{equation}

The previous paragraph describes the construction of a set of triangles $\mathcal{T}=\{r\}$ that is sufficient for our embedding. The next step is to determine the distribution of the couplings of the original problem among the couplings of the individual triangles. Without loss of generality, for each edge $(k, l)$ we set $j_{kl}^{p^*} = J_{kl}$, where $p^*$ indexes the first triangle we encounter containing $(k, l)$, and $j_{kl}^{p \neq p^*} = 0$, where all triangles $\{p\}$ contain $\{k, l\}$. Note that this choice satisfies the condition $\sum_{p / \{k, l\} \subset p} j_{kl}^{p} = J_{kl}$.
\begin{figure}
    \centering
    \includegraphics[width = 0.46\textwidth]{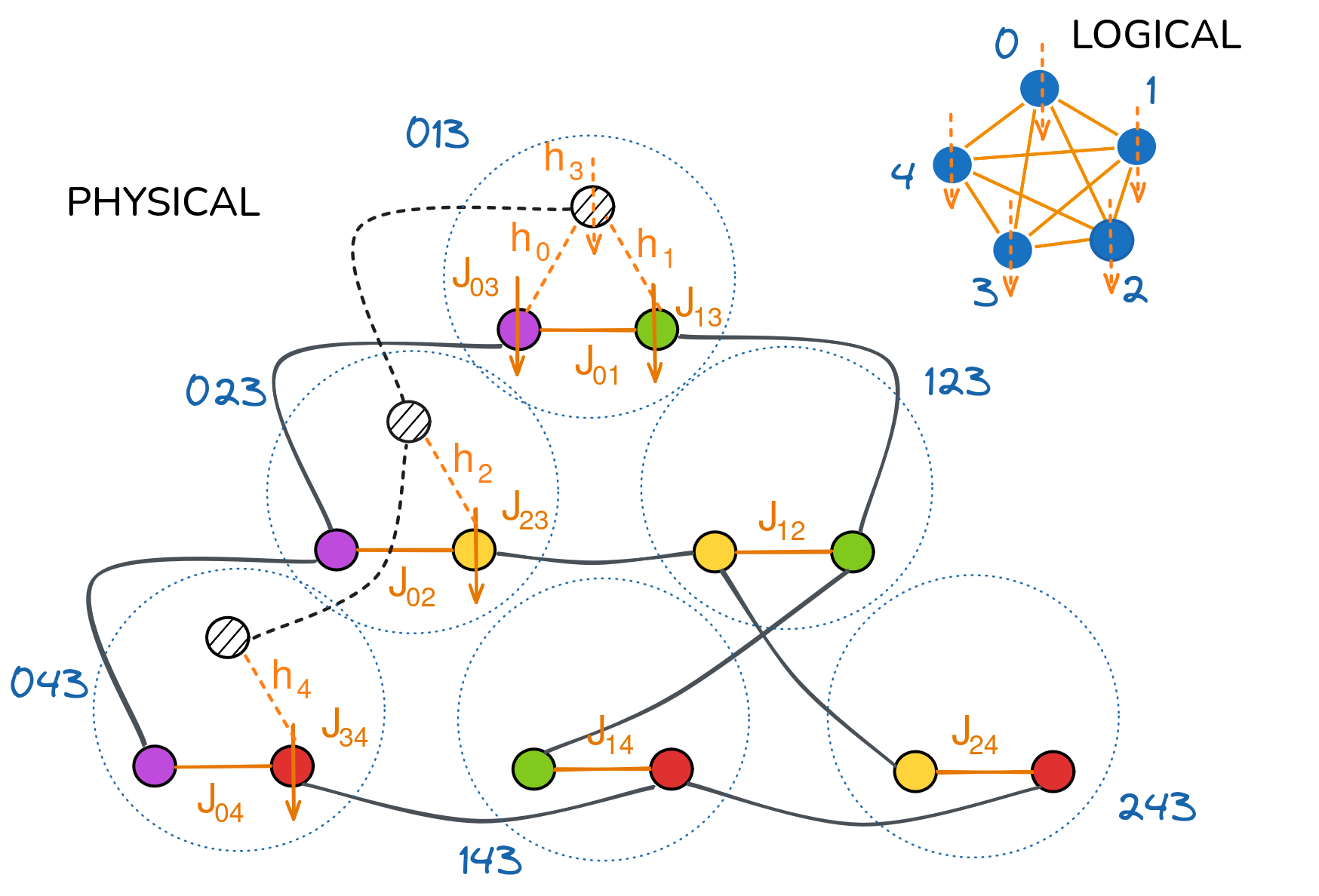}
    \vspace{-0.2cm}
    \caption{Full physical graph (main figure) representing a fully connected logical Ising problem for N=5 (upper right figure) for $k^*=3$. The logical triads $\{r =  \{k, l, m\}\}$ are shown in blue, with blue dotted circles enclosing the physical qubits $\{\eta_r, \xi_r(, \gamma_r)\}$ that encode them. Different qubit colors indicate the ferromagnetic chains formed after imposing the constraints. Sign qubits $\{\gamma_r\}$ (striped) and associated dashed links need not be present in the absence of local fields (i.e., for encoding Eq.~\eqref{eq:ising_Z2}).
    Orange color-coding corresponds to the $\{J_{kl}\}$ (solid) and $\{h_k\}$ (dashed) of the logical problem.}
    \label{fig:full_scheme_N5}
\end{figure}
With this, the procedure to build the full hardware graph can be rephrased as follows. First, we choose the selected node $k^*$, i.e., the logical variable to be involved in all the triangles.
Then, we arrange the pairs of physical qubits representing triangular cells as follows: the first (top-most) row contains triangle $(0, 1, k^*)$, with $k^*$ being the selected node, the second row contains $(0, 2, k^*)$ and $(1, 2, k^*)$, etc., skipping over $k^*$. The resulting graph is illustrated by the solid lines in Fig.~\ref{fig:full_scheme_N5} for the case of $N=5$ and $k^*=3$ (indeed, note that all logical-variable triangles, indicated in the figure by blue triads of numbers, involve the selected node).
After imposing consistency conditions, we end up with strong ferromagnetic links connecting neighboring cells that share an edge in such a way that every physical qubit is connected to at most other 3 in total.
%


%
\subsection{Incorporation of local fields}\label{sec:local_fields}
We can simulate the general Ising model,  Eq.~\eqref{eq:general_ising}, by slightly modifying the current construction: due to the symmetry breaking, now the energies of each triangle $r$ need to be described by three physical qubits, $\{\eta_r, \xi_r, \gamma_r\}$. We refer to the $\{\gamma_r\}$ as sign qubits, as they indicate whether the configuration given by $\{\eta_r, \xi_r\}$ (as defined in the $\mathbb{Z}_2$-symmetric case) has its parallel spins in the 0 or in the 1 state. 
This is illustrated in Table~\ref{tab:frust_configs}b..
The single-cell Hamiltonian $H_r$ giving the right energies is found solving an $8 \times 8$ linear system (we refer again to Appendix~\ref{sec:Hphys_cell} for details).

The consistency conditions regarding sign qubits imply that they are ferromagnetically coupled to the sign qubits of their neighboring cells. However, following the same rule of including the full local field on the first appearance as for the $\mathbb{Z}_2$-symmetric case, the only sign qubits that maintain some connectivity with the rest of their cell are those corresponding to the left-most diagonal. Thus, the remaining sign qubits are irrelevant and can be removed.
The resulting scheme is illustrated for $N=5$ in Fig.~\ref{fig:full_scheme_N5}, where the sign qubits are striped and their associated couplings are dashed.


%
%

\section{Driver Hamiltonians for scalability}\label{sec:drivers_for_scalability}
Two issues arise when embedding a problem within a larger Hilbert space by means of some constraints: (i) the preservation of the logical ground state as the ground state of $H_p$ (which amounts to enforcing the constraints strongly enough) and (ii) the scaling of the minimal gap, since it now decreases with the physical system size $n$ rather than the original logical system size $N$.
The standard driver Hamiltonian for annealing $H_0 = H_{std} = -\sum_i \sx_i$ encodes no information about the feasible solution subspace, and is thus expected to render a search through a space much larger than necessary. This makes the computation less efficient and translates into an exponentially decreasing minimal gap in comparison to the fully connected model. This was also pointed out in the SQA (Simulated Quantum Annealing) study of \cite{albash_simulated-quantum-annealing_2016}, which benchmarked the performance of minor embedding vs. LHZ.
%

However, we do have information about the relevant subspace to explore: it is the space spanned by $\{\ket{0}^{\otimes l}, \ket{1}^{\otimes l} \}$ within each chain, which we will refer to as the logical subspace $\mathcal{L}$. The collection of optimal driver Hamiltonians for the annealing of $H_p$ \footnote{
In the sense of restricting the evolution to the feasible space.
}
should thus commute with the operator $O$ describing the restriction to this subspace \cite{hen_quantum_2016, hen_driver_2016}, i.e., $[O, H_0] = [O, H_p] = 0$, $[H_0, H_p] \neq 0$, and also induce hopping between all regions of the solution space. Notice that for our architecture $O= \sum_{c_k \in \mathcal{C}} \sum_{i\in c_k} \sz_i \sz_{i+1}$, where $\mathcal{C}$ is the set of strongly coupled ferromagnetic chains and $c_k$ indexes each of these. These conditions automatically solve both of the aforementioned issues (i) and (ii) by removing the need for constraints and making the annealing of the embedded problem equivalent to that of the fully connected model.
However, a Hamiltonian of these characteristics for the class of $H_p$ at hand will have highly nonlocal terms \footnote{
Exemplary Hamiltonians of this type can be extracted from a circuit that builds a GHZ state. One such Hamiltonian is:
    \begin{gather}
        H_0 = H_{\text{GHZ}} = - \sum_{c_k \in \mathcal{C}}\left(\prod_{i_r \in c_k} \sx_{i_r} + \sz_{i_0} \sum_{r=1}^{l-1} \sz_{i_r} \right)
        \label{eq:H_GHZ}
    \end{gather}
    where $c_k$ corresponds to the $k$-th chain of the chain set $\mathcal{C}$, all of which are of length $l$. 
    The ground and first excited states of Eq.~\eqref{eq:H_GHZ} correspond to the even and odd GHZ states.
}, rendering its physical implementation impractical.

In order to find more feasible alternatives, we relax the previous conditions such that the commutation of the driver with the constraint is no longer satisfied, but such that the low-energy subspace of the driver has a larger overlap with the logical subspace $\mathcal{L}$.

The easiest driver to implement that fulfills the previous requirement is the ferromagnetic transverse-field Ising model (TFIM):
\begin{gather}
    H_0 = H_{\text{TFIM}} = \sum_{c_k \in \mathcal{C}} h_{\text{TFIM}}^{c_k}
    \label{eq:H_TFIM} \\
    h_{\text{TFIM}}^{c_k} = -\sum_i \sx_i - J_{\text{ZZ}}  \sum_{i \in c_k} \sz_i \sz_{i+1}
    \label{eq:h_TFIM}
\end{gather}
This Hamiltonian is degenerate for $J_{\text{ZZ}} \geq 1$, with the ground subspace spanning the logical space of interest. Notably, $H_{\text{TFIM}}$ preserves the total parity of the system \cite{dziarmaga_dynamics_2005}, as $[h_{\text{TFIM}}^{c_k}, \otimes_{i\in{c_k}}\sx_i] = 0 \ \forall c_k$, which aids the preparation of a state that approaches an even GHZ state by annealing to $H_{\text{TFIM}}$ from the standard initial Hamiltonian $H_{std}$. Note that GHZ $\in \mathcal{L}$ while keeping both configurations, all 0's or all 1's, equally likely. In this manner we can bias our initial ground state towards $\mathcal{L}$ by approaching the ferromagnetic phase as much as possible.
The analytical solution of the transverse Ising chain \cite{lieb_two_1961} (see more modern reviews in \cite{franchini_introduction_2017, mbeng_quantum_2020}) shows that the gap between the even ground state and the first even excited state converges to $\Delta^{\text{TFIM}}_{\infty}(J_{\text{ZZ}}) = 2(1 - J_{\text{ZZ}})$ in the thermodynamic limit in the paramagnetic phase, i.e., for $J_{\text{ZZ}} \leq 1$. 
Thus, the limit on the performance of this driver for large system sizes is given by its gap, since there is a lower bound on the gap for the successful preparation of the ground state of $H_0$. This minimum viable gap is determined by the noise limitations of the annealing device, e.g., thermal excitation.

\begin{figure}
    \includegraphics[width=0.51\textwidth]{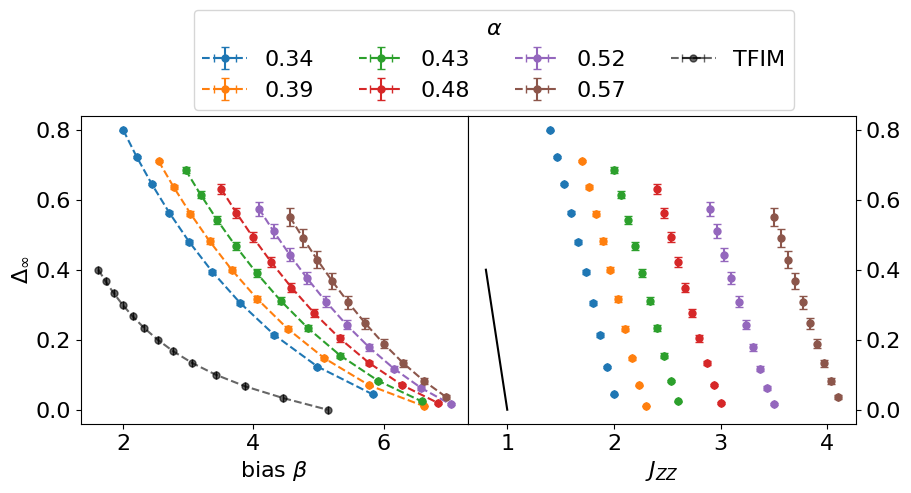}
    \caption{In (a) we present the bias towards $\mathcal{L}$, $\beta$, vs. the gap in the thermodynamic limit $\Delta_{\infty}$ for different values of $J_{\text{ZZ}}$. Error bars on the x-direction are too small to be seen. (b) shows the respective $J_{\text{ZZ}}$ for each $\Delta_{\infty}$.}
    \label{fig:comparison_beta_gap}
\end{figure}

The decay of the projection of the ground state $\ket{GS}$ of \eqref{eq:h_TFIM} onto $\mathcal{L}$ can be fitted to an exponential of the form $\langle P_{\mathcal{L}} \rangle_{GS} = \bra{GS} P_{\mathcal{L}} \ket{GS} = 2^{- l/\zeta + c}$, with $P_{\mathcal{L}} = \ket{0}^{\otimes l}\bra{0}^{\otimes l} + \ket{1}^{\otimes l}\bra{1}^{\otimes l}$ and $l$ the length of the chain. Notice that for $H_{std}$, $\zeta = 1$ and $c=1$. Taking this as the reference for no bias towards the logical subspace, we consider the quantity $\beta = \zeta - 1$ as the bias introduced towards the solution space, such that $H_{std}$ has $\beta=0$. The comparison of $\beta$ vs. $\Delta_{\infty}$ for this model is presented in Fig.~\ref{fig:comparison_beta_gap}, where the $\beta$ values have been extracted from fits of DMRG simulations of up $N=100$.

Following the intuition that the presence of $\sx \sx$ terms will enlarge the gap and thus offer more favorable routes to approach the ferromagnatic phase, we now consider the following driver Hamiltonian:
\begin{gather}
    H_0 = H_{\text{XYZ}} = \sum_{c_k \in \mathcal{C}} h_{\text{XYZ}}^{c_k}
    \label{eq:H_XYZ} \\
    h_{\text{XYZ}}^{c_k} = -\sum_i \sx_i - J_{\text{ZZ}} \sum_{i \in c_k} (\sz_i \sz_{i+1} + \alpha \sx_i \sx_{i+1})
    \label{eq:h_XYZ}
\end{gather}
The previous driver is nothing but the XYZ model. In the absence of local fields it is integrable~\cite{baxter_exactly_1990} and has been shown to present a ferromagnetic phase in the $z$-direction for $\alpha < 1$~\cite{shi_duality_2020}. The fixed local field in the x-direction present in Eq.~\eqref{eq:h_XYZ} shifts the boundary of this phase to lower $\alpha$ values at fixed $J_{\text{ZZ}}$ or, equivalently, to higher $J_{\text{ZZ}}$ as $\alpha \to 1$.
We numerically study this model with DMRG in order to evaluate $\beta(J_{\text{ZZ}}, \alpha)$ and find the gap in the thermodynamic limit. For the latter purpose we consider the following fitting function: $\Delta(l) = A l^{-\Omega} + \Delta_{\infty}$, where the algebraic decay corresponds to the behavior of the gap near the phase transition. This functional form accurately describes the gap of the finite-sized TFIM, so one can expect a reasonably good agreement with the true functional form at least for low $\alpha$. We highlight that these fits are performed for each $(J_{\text{ZZ}}, \alpha)$ point, and refer to Appendix~\ref{sec:numerical_methods} for details. Fig.~\ref{fig:comparison_beta_gap} shows how appropriate combinations ($J_{\text{ZZ}}, \alpha$) can further increase $\beta$ with respect to the TFIM driver. As shown, tuning the relative strength of the XX couplings allows to achieve a greater bias towards $\mathcal{L}$ for equal gap sizes, at the cost of increasing the energy scale.

\begin{tikzpicture} [remember picture, overlay]
        \draw ([xshift=15.3cm,yshift=-3.4cm]current page.north west) circle [radius=0cm] node{a)};
        \draw ([xshift=19.2cm,yshift=-3.4cm]current page.north west) circle [radius=0cm] node{b)};
\end{tikzpicture}

%
\section{Discussion}\label{sec:discussion}
We have derived a 2-local hardware graph of minimal connectivity that is sufficient to encode any arbitrary Ising problem. Indeed, a graph of degree $d=3$ is the smallest to hold complex problems, such as the NP-complete SAT problem \cite{barahona_computational_1982}.
There is a dual interpretation of the resulting graph: in terms of consistent triangular cells and in terms of chains denoting logical variables. The full architecture, incorporating local fields, requires $n = N(N-2)$ physical qubits, and
the sparsity of the graph implies less crosstalk to compensate for, an important scalability feature.
Achieving strong couplings without compromising coherence is a challenging task, but one that needs to be tackled nonetheless for any embedding strategy that enables scalability. In the particular case of superconducting technology, already standard couplers naturally provide a much larger ferromagnetic range than the antiferromagnetic one \cite{weber_coherent_2017}. We have contributed to this scalability challenge from the algorithmic side by proposing alternative driver Hamiltonians that enhance the baseline success probability of the anneal by directing the search towards the logical subspace, which mitigates the need for strong penalties. Some of these Hamiltonians can be readily implemented with current devices.


The proposed scheme has several important advantages: the embedding of the problem does not require any additional costly pre-computations, the maximal sparsity of the graph aids noise and crosstalk minimization and the interactions required are within the reach of current technology. In addition, the strongly coupled chains of qubits that compose this architecture allow to spatially spread the information, making the computation more robust to some external error sources. Finally, the structure of the logical space allows to devise relatively simple driver Hamiltonians that mitigate the demands for strong constraint terms. In future work we will focus on studying the extent of this mitigation in order to assess the feasibility of implementing the proposed architecture in real hardware.

\appendix

\section{ON THE IMPLICIT CHANGE OF VARIABLES}\label{sec:factorised_encoding}

In Fig. 1 of the main text it can be noticed that the consistency conditions lead to a network of $N$ chains of qubits encoding the following logical variables: the orientation of the selected node and the parities of the rest of nodes with respect to the latter. 
Indeed, the collapse of all strong ferromagnetic links into single qubits leads to another all-to-all connected Ising model, if the original one had full connectivity as well.
Thus, taking only the structure of these logical chains the triangular cell interpretation could be discarded and one could directly encode the original problem variables, i.e., the orientation of each spin. However, the triangle description naturally leads to a transformation of the original problem that deserves attention of its own, since it can be beneficial in some cases. 
This transformation was already pointed out in \cite{fujii_energy_2022, fujii_eigenvalue-invariant_2023}, where it was referred to as ELTIP (Energy Landscape Transformation of Ising Problem) and numerically studied for small instances. We will hereby provide a general overview of its potential in broader terms.

We will refer to this kind of transformation as mixed encodings, referencing the fact that it mixes parity with single-site variables. Taking $k^*=0$ without loss of generality, this transformation corresponds to the variable change $x_0 \to \eta_0$, $x_{j\neq 0} x_0 \to \eta_j$, such that $H(\vec{x}) = \sum_i h_i x_i + \sum_{j>i} J_{ij} x_i x_j  \to H(\vec{\eta})$ with
\begin{gather}
    H(\vec{\eta}) = 
    h_0 \eta_0 + 
    \sum_{i>0}^{N-1} J_{i0} \eta_i +
    \sum_{i>0}^{N-1}  h_i \eta_i \eta_0  + \sum_{i>0}^{N-1} \sum_{j > i}^{N-1} J_{ij} \eta_i \eta_j
    \label{eq:variable_transformation}
\end{gather}
where $N$ is the size of the problem. This new choice of logical variables induces a reordering of the computational basis states, which may lead to smaller Hamming distances $\mathcal{D}(m, n)$
\begin{equation}
    \mathcal{D}(m, n) = \bra{m} \sum_{i=0}^{N-1} \sx_i \ket{n}
    \label{eq:Hamming_distance}
\end{equation}
between ground- and lower-excited states, respectively referred to GS and $m$-EXC from now on, and thus the mitigation of perturbative crossings \cite{amin_first-order_2009, altshuler_anderson_2010, farhi_quantum_2011, albash_adiabatic_2018} at the end of the anneal. Note that we refer to $m$-excited states in general, despite the fact that the anticrossing will always take place between the GS and instantaneous 1-EXC states, where the $m$ is the eigenlevel of the final Hamiltonian the respective instantaneous 1-EXC is connected to. We do this to account for the cases in which there are several anticrossings.

It is worth noting that the worse the perturbative crossing (i.e., the smaller the gap it induces), the larger the Hamming distance between participating GS and $m$-EXC. In order for the transformation of Eq.~\eqref{eq:variable_transformation} to modify a given configuration, the selected qubit $k^*$ must be in whichever state we consider as denoting antiparallel spins in the mixed picture. Therefore, only pairs of configurations in which the selected qubit has opposite orientations will have their Hamming distance modified: if two states have the same configuration in $k^*$, either they don't change at all or they both flip the rest of qubits $k\neq k^*$, which in either scenario leads to the same Hamming distance. In the case of opposite orientations in $k^*$, two bitstrings $\phi_\alpha, \phi_\beta$ corresponding to levels $\alpha, \beta$ with an original distance $\mathcal{D}(\phi_\alpha, \phi_\beta) = N-q$ for some $q \in \{0, 1, 2, ..., N-1\}$ are transformed such that $\mathcal{D}(\Tilde{\phi}_\alpha, \Tilde{\phi}_\beta) = q+1$.
However, an improvement will generally not be as drastic as what could be expected for a single relevant excited state, since a typical hard instance for quantum annealing will have a large collection of local minima. In any case, one can still efficiently try to find some improvement over the landscape generated by the original problem, since the family of mixed encoding formulations grows only linearly with system size. 

\begin{figure}
    \centering
    \includegraphics[width=0.7\linewidth]{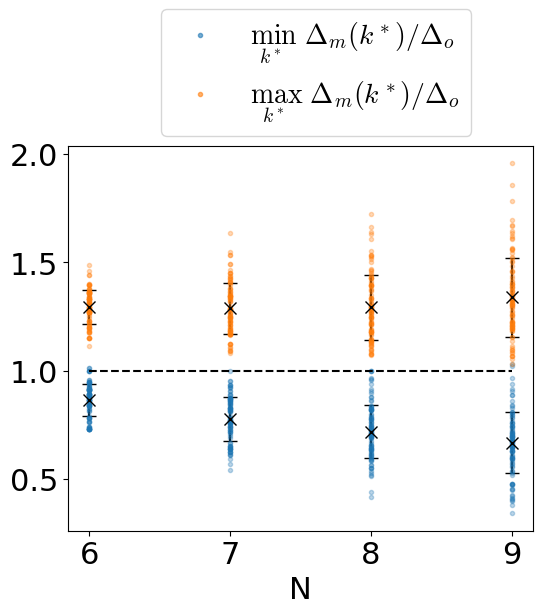}
    \caption{Best- (orange) and worst-case (blue) scenarios of the ratio between the the minimal gaps in the mixed encoding and original problem formulations for a standard quantum annealing protocol. The problem under consideration is UNIQUE 3-SAT. Colored markers correspond to single realizations and the black crosses and error bars indicate the mean and standard deviation. The dashed line has been added to help visualization.}
    \label{fig:mixed_encodings}
\end{figure}

We briefly illustrate the effect of considering mixed encodings for the UNIQUE 3-SAT problem in small system sizes by analyzing the minimum gap of standard annealing processes with a transverse-field driver and linear schedule, averaging over a 100 randomly generated instances. We emphasize that this problem does not have any particular $\mathbb{Z}_2$-symmetric features, so no drastic improvement of the gap is expected due to its reformulation in terms of a mixed encoding. However, this situation is more representative of a generic use-case. We present the results of this study in Fig.~\ref{fig:mixed_encodings}, where it is observed that the average maximum improvement of the minimal gap $\Delta_m$ over that of the original formulation $\Delta_o$ for this particular problem is around 1.3 and appears independent of system size. The worst-case scenario, on the other hand, does seem to provide increasingly small minimum gaps as we scale up in system size. However, reasonable heuristics can be developed in order to guide the set of $\{k^*\}$ to try out, which can be determined by parameters like resulting average coupling strength (vs. resulting average local fields) or resulting coupling strength diversity.

\section{DETERMINING THE PHYSICAL HAMILTONIAN THAT DESCRIBES A LOGICAL TRIANGLE}\label{sec:Hphys_cell}
In this section we detail the step that goes from the Hamiltonian of an individual logical triangle, $H_{r}$, to the physical Hamiltonian that describes it, $H^{r}_{\text{physical}}$. 

\begin{figure}
    \centering
    \includegraphics[width=0.99\linewidth]{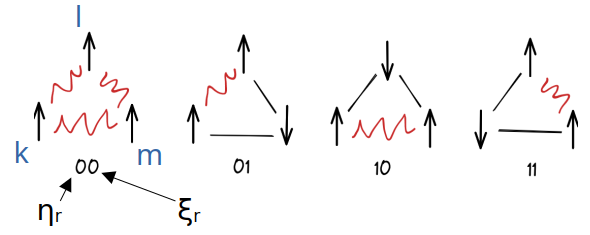}
    \caption{Illustration of the mapping between the configuration of physical qubits $\{\eta_r, \xi_r\}$ and the energetically distinct configurations of the logical triangle $r=\{k, l, m\}$ in the $\mathbb{Z}_2$-symmetric case.}
    \label{fig:internal_labelling_triangle}
\end{figure}

In the $\mathbb{Z}_2$-symmetric case, the relevant logical variables $r=\{k, l, m\}$ are described by two physical qubits alone, $\{\eta_r, \xi_r\}$.
\begin{gather}
    H_{r} = 
        j^{r}_{kl} \sz_k \sz_l + 
        j^{r}_{km} \sz_k \sz_m + 
        j^{r}_{lm} \sz_l \sz_m \\
    H^{r}_{\text{physical}} = 
        \Tilde{h}_{\eta_r} \sz_{\eta_r} + 
        \Tilde{h}_{\xi_r} \sz_{\xi_r} + 
        \Tilde{J}_{\eta_r \xi_r} \sz_{\eta_r} \sz_{\xi_r}
\end{gather}
At this stage, the logical triangle couplings $\{j^{r}_{kl}\}$ are known and what we whish to determine are the variables $\Tilde{h}_{\eta_r}, \Tilde{h}_{\xi_r}$ and $\Tilde{J}_{\eta_r \xi_r}$.
In order to do this, we first establish the internal labeling of triangle $r$, i.e., the mapping between the (energetically distinct) logical configurations and the configuration of the physical qubits $\eta_r, \xi_r$. An illustration of this mapping is presented in Fig.~\ref{fig:internal_labelling_triangle}. As mentioned in the main text, we assign the physical ``00" configuration to the all-parallel logical configuration. The remaining assignments are free as long as the ``11" configuration does not indicate parallel spins in an edge shared with another logical triangle. We recall that this condition is necessary in order to have exclusively 2-local constraints arising from the consistency conditions among triangles.

Once the assignment is fixed, we can determine the energies of the individual configurations $\{e_{\eta_r \xi_r}\}$ and thus obtain the couplings and local fields necessary to generate them:
\begin{equation}
    \begin{pmatrix}
        1 & 1 & 1 & 1 \\ 
        1 & 1 & -1 & -1 \\ 
        1 & -1 & 1 & -1 \\ 
        1 & -1 & -1 & 1
        \end{pmatrix}
        \begin{pmatrix}
            \alpha \\ \tilde{h}_{\eta_r} \\ \tilde{h}_{\xi_r} \\ \tilde{J}_{\eta_r \xi_r}
        \end{pmatrix} = 
        \begin{pmatrix}
            e_{00} = j_{kl} + j_{km} + j_{lm} \\
            e_{01} = j_{kl} - j_{km} - j_{lm} \\
            e_{10} = - j_{kl} + j_{km} - j_{lm} \\
            e_{11} = - j_{kl} - j_{km} + j_{lm}
        \end{pmatrix}\, .
\end{equation}
Notice that the coefficient matrix on the left-hand side of the previous equation is built from the diagonals of the operators $I, \sz_{\eta_r}, \sz_{\xi_r}$ and $\sz_{\eta_r} \sz_{\xi_r}$ as columns.
In this manner, we find that
\begin{equation}
    \alpha = 0, \qquad \tilde{h}_{\eta_r} = j_{kl}, \qquad \tilde{h}_{\xi_r} = j_{km}, \qquad \tilde{J}_{\eta_r \xi_r} = j_{lm}\, , 
\end{equation}
as stated in the main text.

For a general Ising model (i.e., with local fields) the procedure is completely analogous: determine the internal labeling of the logical triangle (this time with respect to $\xi_r, \eta_r$ and the sign qubit $\gamma_r$); build the coefficient matrix from the diagonals of the operators $I, \sz_{\eta_r}, \sz_{\xi_r}, \sz_{\gamma_r}, \sz_{\eta_r} \sz_{\xi_r}, \sz_{\eta_r} \sz_{\gamma_r}$ and $\sz_{\xi_r} \sz_{\gamma_r}$; and solve for $\alpha, \tilde{h}_{\eta_r}, \tilde{h}_{\xi_r}, \tilde{h}_{\gamma_r}, \tilde{J}_{\eta_r \xi_r}, \tilde{J}_{\eta_r \gamma_r}$ and $\tilde{J}_{\xi_r \gamma_r}$.

\section{NUMERICAL METHODS}\label{sec:numerical_methods}

Numerical DMRG simulations of the 1D chains were performed using the library ITensor \cite{itensor}. 
The maximum bond dimension for the simulations was $\chi=500$, well above the maximum bond dimension observed in the converged results, and the number of sweeps was 30 for every point. For each $(J_{\text{ZZ}}, \alpha)$ pair, systems sizes of from $N=10$ to $N=100$ were simulated in steps of 10. 

\begin{figure*}
    \centering
    \includegraphics[width=0.65\textwidth]{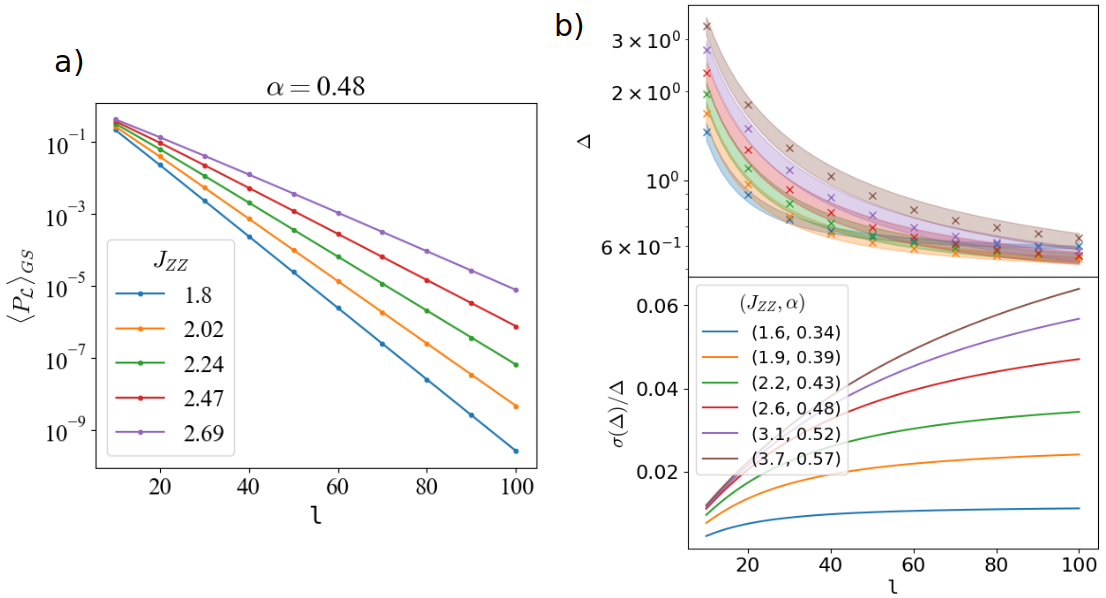}
    \caption{Illustration of the accuracy of the fits for some exemplary values of $(J_{ZZ}, \alpha)$. In (a) we present the exponential decay of the projection of the ground state onto the logical subspace. In (b) we present the $(J_{ZZ}, \alpha)$ data corresponding to the largest error in the fit to Eq.~\eqref{eq:fit_gap_vs_l} per $\alpha$ value, which coincide with the lowest $J_{\text{ZZ}}$ (i.e., furthest from the transition point). The top plot shows the numerically obtained gaps and the resulting fits, where the shaded region corresponds to one standard deviation as derived from the covariance matrix returned by the optimizer. On the bottom plot, the relative error calculated from the covariance matrix returned by the optimizer is presented.}
    \label{fig:illustrative_fit_accuracy}
\end{figure*}

\begin{figure*}
    \centering
    \includegraphics[width=0.7\textwidth]{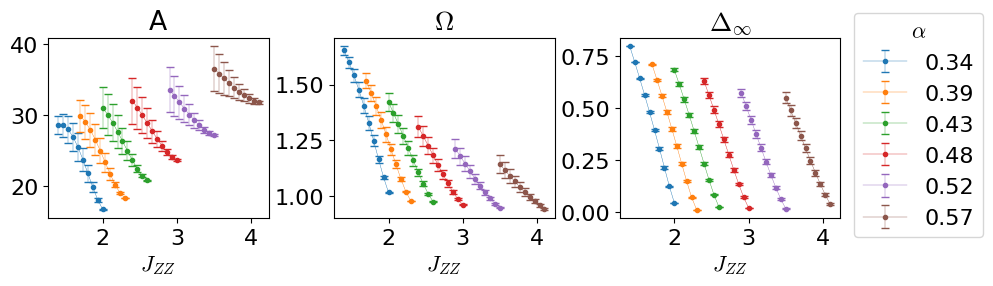}
    \caption{All resulting parameters of the fits to Eq.~\eqref{eq:fit_gap_vs_l}. $\Delta_{\infty}$ (right-most plot) is the sought-after parameter for this work, and is presented in Fig. 2 of the main text. The error bars are calculated from the covariance matrix returned by the optimizer.}
    \label{fig:params_gap_extrapolation}
\end{figure*}

We hereby justify the functional dependencies discussed in the main text. 
We begin by illustrating the decay of the projection of the ground state onto $\mathcal{L}$ with system size in Fig.~\ref{fig:illustrative_fit_accuracy}a, which as expected is exponential due to the exponential increase of the Hilbert space with system size, while the dimension of the logical subspace $\mathcal{L}$ remains constant. The decay is slower for the higher values of $J_{\text{ZZ}}$ in Fig.~\ref{fig:illustrative_fit_accuracy}a, since as we approach the critical $J_{\text{ZZ}}$ (around $J_{\text{ZZ}}=3$, see right-most plot in Fig.~\ref{fig:params_gap_extrapolation}) the eigenstates are more similar to those of the ferromagnetic phase.


We now discuss the fitting function employed for the extrapolation of the gap of the $XYZ$ chain to the thermodynamic limit
\begin{equation}
    \Delta(l) = A l^{-\Omega} + \Delta_{\infty}
    \label{eq:fit_gap_vs_l}
\end{equation}
As discussed in the main text, the above expression is motivated by the algebraic decay that we expect near a second order phase transition. The fit was performed using scipy's optimizer \textit{curvefit}. Fig.~\ref{fig:illustrative_fit_accuracy}b presents some illustrative fits focused on the region of interest, where the agreement between the the fitted curves and the data can be observed. As a general trend, illustrated in Fig.~\ref{fig:params_gap_extrapolation}, the fitting function becomes increasingly less accurate as we go further away from the critical point, and this effect becomes more prominent for higher $\alpha$. Since the ansatz in~\eqref{eq:fit_gap_vs_l} was formulated based on the known solution for the TFIM ($\alpha=0$), it is no surprise that other finite size effects could appear for $\alpha>0$, becoming increasingly important as we go away from the TFIM limit. However, these unaccounted-for effects are, in the end, overridden by the polynomial decay of the gap as we approach the second-order transition. Despite the higher uncertainties in $A$ and $\Omega$ away from the critical point, the optimizer returns a low uncertainty of the extrapolation of the gap in the thermodynamic limit in all cases.

%
\section*{Acknowledgments}
We thank Ramiro Sagastizabal, Matthias Werner, Arnau Riera and Josep Bosch for valuable discussions. This work was supported by European Commission EIC-Transition project RoCCQeT (GA 101112839), the Agencia de Gestió d'Ajuts Universitaris i de Recerca through the DI grant (No. DI74) and the Spanish Ministry of Science and Innovation through the DI grant (No. DIN2020-011168). This work has been funded by Grant PID2023-147475NB-I00 funded by MICIU/AEI/10.13039/501100011033 and FEDER, UE, by Grant 
No. 2021SGR01095 from Generalitat de Catalunya, and by 
Project CEX2019-000918-M of ICCUB (Unidad de Excelencia María 
de Maeztu).

\bibliography{references}

\end{document}